# How to couple identical ring oscillators to get Quasiperiodicity, extended Chaos, Multistability and the loss of Symmetry.

Edward H. Hellen[a], Evgeny Volkov[b]*

[a] Department of Physics and Astronomy, University of North Carolina Greensboro, Greensboro, NC USA

[b] Department of Theoretical Physics, Lebedev Physical Institute, Leninsky 53, 119991 Moscow, Russia

* - corresponding author, volkov@lpi.ru

**Abstract**

We study the dynamical regimes demonstrated by a pair of identical 3-element ring oscillators (reduced version of synthetic 3-gene genetic Repressilator) coupled using the design of the 'quorum sensing (QS)' process natural for interbacterial communications. In this work QS is implemented as an additional network incorporating elements of the ring as both the source and the activation target of the fast diffusion QS signal. This version of indirect nonlinear coupling, in cooperation with the reasonable extension of the parameters which control properties of the isolated oscillators, exhibits the formation of a very rich array of attractors. Using a parameter-space defined by the individual oscillator amplitude and the coupling strength, we found the extended area of parameter-space where the identical oscillators demonstrate quasiperiodicity, which evolves to chaos via the period doubling of either resonant limit cycles or complex antiphase symmetric limit cycles with five winding numbers. The symmetric chaos extends over large parameter areas up to its loss of stability, followed by a system transition to an unexpected mode: an asymmetric limit cycle with a winding number of 1:2. In turn, after long evolution across the parameter-space, this cycle demonstrates a period doubling cascade which restores the symmetry of dynamics by formation of symmetric chaos, which nevertheless preserves the memory of the asymmetric limit cycles in the form of stochastic alternating "polarization" of the time series. All stable attractors coexist with some others, forming remarkable and complex multistability including the coexistence of torus and limit cycles, chaos and regular attractors, symmetric and asymmetric regimes. We traced the paths and bifurcations leading to all areas of chaos, and presented a detailed map of all transformations of the dynamics.

# Highlights

- Two identical genetic Repressilators are coupled via additional network borrowed from quorum sensing mechanism of bacterial communications.

- Two-frequency torus, complex anti-phase limit cycles and chaos are found over very large areas of control parameters.

- Asymmetrical limit cycle coexisting with symmetric chaos is discovered and its evolution is studied.

- The paths to chaos across the regions with multistability are presented.

## 1. Introduction

Coupled oscillators are the most frequently used model to study such collective phenomena as synchronization [1], wave generation, multistability [2] etc. in all fields of fundamental science and its applications. Coupling can quench oscillations by either suppressing all oscillators to the same fixed point (amplitude death) or by creating an inhomogeneous stable steady-state (oscillation death, see [3] for review) which shares the phase space with other oscillatory regimes. There are many designs of coupling initiated by studies of real systems: starting from the old classic observations of pulse-coupled fireflies, direct scalar or vector reagents exchange between chemical reactors, electric and/or synaptic interactions between neurons, up to very recent investigations of combined coupling between chemical oscillators [4], half-center oscillator configurations constituted by two bursters [5], and plant interactions [6], which attempt to explain biennial rhythm in fruit production. Nonlocal connections between oscillators can lead to the formation of chimeras (coexistence of coherent and incoherent clusters) in homogeneous populations, even of very different natures: phase oscillators [7, 8], chemical oscillators [9], metronome ensemble [10], see [11] for review. However, chimera death is also determined by the particular coupling mechanism [12].

During the last decade new experimental objects – synthetic genetic oscillators – have attracted great attention [see e.g. reviews: [13, 14] as a new tool for probing the mechanisms of gene expression regulation and as a possible instrument for genetic therapy. We explore the ring-type oscillator which is a well-known circuit in physics and applied technology [e.g. review [15]], and became very popular in synthetic biology after 2000 under the name of "Repressilator" [16] due to its actual assembly and insertion into the bacterial cell E.coli. The Repressilator consists of three genes whose protein products (A, B, C) repress the transcriptions of each other unidirectionally in a cyclic way (..A--┤B--┤C--┤A..). Recently this circuit has been improved upon [17, 18], "making it an exceptional precise biological clock" [19]. The electronic analog of the Repressilator was presented in [20, 21]. The cooperativity of transcription repression, which is the core process of the Repressilator, is typically described by the Hill function $\sim \alpha/(1+x^n)$ where **x** is repressor abundance. The dynamics of coupled Repressilators is very sensitive to the value of **n** which controls the steepness of repression and **α** which determines the amplitude of the isolated Repressilator.

The effectiveness of a genetic oscillator such as the Repressilator depends on how they can function collectively, thereby requiring a coupling method. An almost obvious suggestion was to use the natural bacterial quorum sensing (QS) mechanism, used for cell-cell communication in bacterial populations [22, 23], as the instrument to synchronize genetic oscillators located in different cells. The core of QS is the production of small molecules (autoinducer) which can, first, easily diffuse across the cell membrane and external medium and, second, work to activate/repress transcription of an intended target gene. By manipulating the positions of the gene providing the autoinducer production and QS-sensitive promoters controlling the transcription of other genes in the genetic circuits, one can obtain different coupling types and, as a result, different sets of collective modes in populations of synthetic genetic oscillators. If the autoinducer is not only a signal molecule but also works as an integral participant of the oscillator then its diffusion provides the direct coupling which supports, for example, the wave propagation in bacterial populations [24].

Alternatively, QS may be implemented in a genetic oscillator as an additional element not required for the generation of the auto-oscillations. In this case the coupling may be described as indirect because its activity is mediated by the complex chain: production of autoinducer, its diffusion and binding with the promoter of the target gene. Such coupling is typically nonlinear because of bimolecular interaction of autoinducer and target instead of the linear intercellular diffusion of similar variables in direct coupling. This coupling method has been explored in model simulations of QS-coupled relaxation oscillators [25, 26] and repressilators [27], and its ability to synchronize the in-phase regime for detuned populations was demonstrated. Further work [28] has demonstrated that the model [27] may exhibit more complex collective regimes if the ranges of key parameters are reasonably extended. Later, in-phase synchronization of coupled Repressilators in the presence of noise was demonstrated [29]. More complex and flexible dynamics, which better corresponds to biological diversity, have been demonstrated in [30, 31].

In this work we couple two Repressilators via the scheme of quorum sensing, as suggested in [32, 33]. In this version of coupling it is important that the production of signal molecules is associated with the gene which is located inside the Repressilator ring upstream with respect to the target gene, which accepts the impact of autoinducer. This means that this coupling should be classified as "conjugate", because in the ordinary differential equation (ODE) system (1) (see the next section), the production of the autoinducer and its effect on the target gene are controlled by the equations for different repressors. Intercellular communication is realized by the simple diffusion of autoinducer but the coupling as a whole is more complex since it cannot be reduced to the well-studied diffusion-controlled ODE. Such a coupling seems quite acceptable for real genetic networks and nevertheless may be constructed in artificial networks of a different nature.

Recently [34] we have demonstrated by numerical and electronic simulations that two coupled identical Repressilators show the development of very flexible dynamics if the **n** increases up to **n** = 4. We found spatially homogeneous and inhomogeneous types of chaos over large areas of the coupling strength (**Q**) as well as the set of periodic windows which contains inhomogeneous limit cycles partially synchronized with i:j winding number, e.g. i=1,2…, j=i+1 despite the identical nature of the oscillators.

Our main goal here is to trace the routes from self-organized quasiperiodicity to the unusual collective regimes, including homogeneous chaotic regimes and inhomogeneous limit cycles. We present the detailed map of isolated and overlapping regimes over a large plane of the key parameters **Q** and **α** for other parameters fixed. It is known that in the model [33] the anti-phase (**AP**) limit cycle, which is the single stable attractor under small coupling strength, loses stability via torus bifurcation if coupling strength increases. The boundaries of stable self-organized quasiperiodic regimes delineate a very large region in the **Q**-**α** plane inside which many dynamic regimes exist. To present the results in a foreseeable form, we restrict the ratio **α**-max / **α**-min to 25, based on a reasonable scatter of the corresponding experimental data on the difference between weak and strong promoters [35, 36].

Two main scenarios of dynamics inside the (**Q**-**α**)-plane after torus creation will be presented. The first one is realized for the values of **α** just above **α**-min, defined to be the smallest **α** for which the AP limit cycle loses stable continuity during variation of coupling strength **Q**. Here, the evolution of the torus progresses towards chaos but encounters a complex symmetric limit cycle with 5 return times (LC5:5) in each Repressilator. The torus and the LC5:5 have similar structures, they co-exist, and after destruction of the torus, the system sits on the limit cycle, which demonstrates the standard period doubling cascade, leading to chaos if **Q** and/or **α** increase. This symmetrical LC5:5 is an isolated attractor with unclear bifurcation origin and is unrelated to standard "resonant" cycles on the torus.

The second scenario starts for larger **α** where the LC5:5 loses stability at smaller **Q**-values than does the torus, and torus destruction, rather than period doubling of LC5:5, provides the chaotic oscillations. Increasing **Q** in the **Q**-**α** plane, chaos meets the stability boundary for the spatially asymmetric (inhomogeneous) LC1:2, after which both attractors coexist in a wide (**Q**-**α**)-band. Further, at the upper boundary of this band the chaos becomes unstable, converting to the inhomogeneous LC1:2 which develops alone demonstrating both the period doubling cascade, ending in spatially homogeneous (symmetric) chaos with very "polarized" time series and torus bifurcation resulting in inhomogeneous quasiperiodic regimes.

Both scenarios are investigated in detail by numerical bifurcation analysis as well as by direct calculations of time series and Lyapunov exponents. Although we used identical and fairly smooth oscillators, the proposed QS-coupling scheme generated an unusual cascade of complex homogeneous and inhomogeneous solutions coexisting in many areas of the phase diagram.

## 2. Model and Methods

We investigate the dynamics of two Repressilators interacting via repressive QS coupling as used previously [32, 33]. Figure 1 shows two repressilators located in different cells and coupled via QS to the external medium. The three genes in the loop produce mRNAs (a, b, c) and proteins (A, B, C), and they impose Hill function inhibition on each other in cyclic order by the preceding gene. The QS feedback is maintained by the AI produced (rate $k_{S1}$) by the protein B while the autoinducer (AI) communicates with the external environment and activates (rate κ in combination with Michaelis function) production of mRNA for protein C, which, in turn, reduces the concentration of protein A resulting in activation of protein B production. In this way the protein B plays a dual role of direct inhibition of protein C synthesis and AI-dependent activation of protein C synthesis, resulting in complex dynamics of the repressilator, even for just a single repressilator [37].

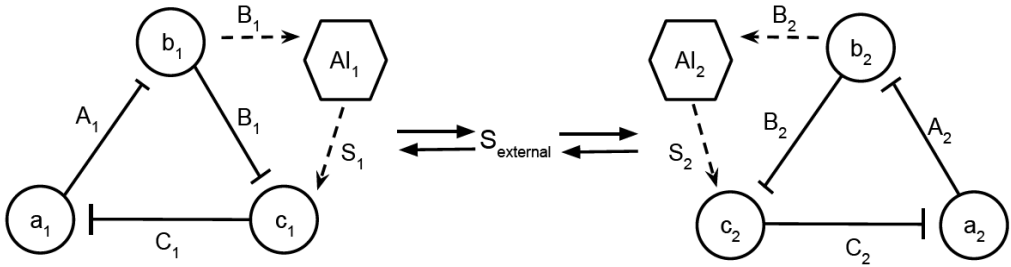

**Fig.1.** A genetic network from two repressilators with QS feedback and the global coupling due to the mixing of signal molecules in extracellular medium. Lower case (a, b, c) are mRNAs and upper case (A, B, C) are expressed protein repressors. $S_i$ is the autoinducer molecule which diffuses through the cell membrane.

The original models of a single repressilator [16, 33] used re-scaled dimensionless quantities for rate constants and concentrations. We reduce the model for the case of fast mRNA kinetics ((a, b, c) are assumed in steady state with their respective inhibitors (C, A, B), so that $da/dt = db/dt = dc/dt \approx 0$). The resulting equations for the protein concentrations and AI concentration S are,

$$\frac{dA_i}{dt} = \beta_1\left(-A_i + \frac{\alpha}{1+C_i^n}\right) \qquad (1a)$$

$$\frac{dB_i}{dt} = \beta_2\left(-B_i + \frac{\alpha}{1+A_i^n}\right) \qquad (1b)$$

$$\frac{dC_i}{dt} = \beta_3\left(-C_i + \frac{\alpha}{1+B_i^n} + \frac{\kappa S_i}{1+S_i}\right) \qquad (1c)$$

$$\frac{dS_i}{dt} = -k_{S0}S_i + k_{S1}B_i - \eta(S_i - S_{ext}) \qquad (1d)$$

where i = 1,2 for the two repressilators, $\beta_j$ (j = 1,2,3) are the ratios of protein decay rate to mRNA decay rate, α accounts for the maximum transcription rate in the absence of an inhibitor, and **n** is the Hill cooperativity coefficient for inhibition. For the quorum sensing pathway $k_{S0}$ is the ratio of the AI decay rate to mRNA decay rate, and as previously mentioned, $k_{S1}$ is the rate of production of AI and κ gives the strength of AI activation of protein C. The diffusion coefficient η depends on the permeability of the membrane to the AI molecule. The concentration of AI in the external medium is $S_{ext}$ and is determined according to quasi-steady-state approximation by AI produced by both repressilators ($S_1$ and $S_2$), and a dilution factor **Q**.

$$S_{ext} = Q \frac{S_1 + S_2}{2} \qquad (2)$$

Numerical simulations are performed with XPPAUT [38], AUTO-07p [39], and by direct integration with 4$^{th}$-order Runge-Kutta solver. We choose parameter values similar to ones used previously. Here we use $\beta_1 = 0.5$, $\beta_2 = \beta_3 = 0.1$, n = 3.0, k=15, $k_{S0} = 1$, $k_{S1} = 0.01$, and η = 2. Parameters **α** and **Q** are chosen for bifurcation analysis as they represent the amplitude of the individual oscillators, and the strength of the coupling.

## 3. Results

### *3.1 Low-strength Oscillators*: **α < 1000**

To start we present the basic coarse-grained map of regular oscillating regimes and steady states in system (1), which may coexist with new regimes discussed below or be located in the parameter space around them. One-parameter XPPAUT continuations shown in Fig. 2 reveal three branches; a homogeneous steady state (HSS) spanning the Q-range, an inhomogeneous SS (IHSS) arising from pitchfork bifurcations of HSS (BP1 and BP2), and an anti-phase LC (APLC) arising from the Andronov-Hopf bifurcation (HB). The steady-states have stable (red) and unstable (thin black) sections, with the stable IHSS occurring between limit points LP2 and LP3. There is a narrow range between HB and LP4 of stable low level HSS which coexists with the stable high value HSS. The APLC is stable (green) at low and high Q-values, and is unstable (thin blue) between the torus bifurcations of APLC (TR1 and TR2). The APLC is the sole stable attractor at low Q (see [32] for details).

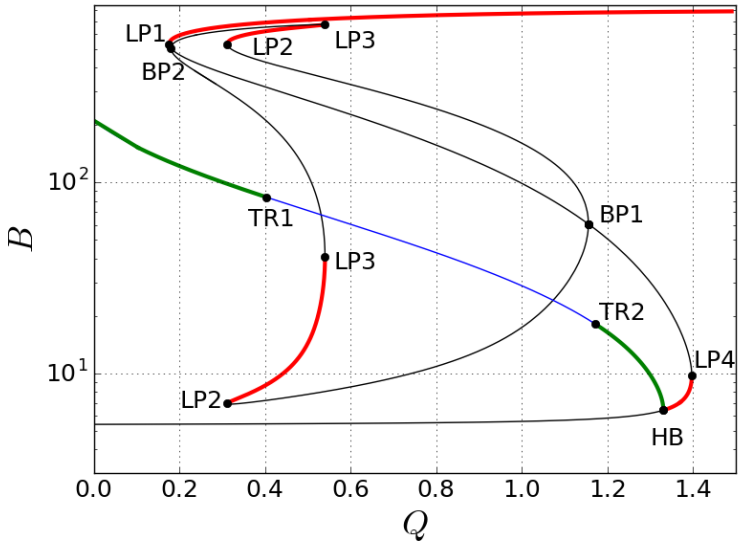

**FIG. 2.** The amplitudes of APLC (green line) and HSS, IHSS steady states (red lines) as function of **Q**. Solid (thin) lines are stable (unstable) solutions. TR, LP, BP, HB are designations of torus, limit point, branch point (pitchfork), Andronov-Hopf bifurcations, respectively. n=3, k=15, α=800. Other parameters are fixed: $\beta_1 = 0.5$, $\beta_2 = \beta_3 = 0.1$, n = 3.0, $k_{S0} = 1$, $k_{S1} = 0.01$, and $\eta = 2$

The main goal of the current work is the study of dynamics inside the region with unstable APLC. Figure 2 shows that this area is large for the given parameters. Its size depends on the values of other model parameters but it will be large throughout our work and the structure of the phase diagram in Fig.2 will not be changed qualitatively. Any new attractors share the phase space with HSS and IHSS but the basins of steady states are not so large as to suppress the exhibition of new regimes.

We choose **n=3, k=15** to study the evolution of dynamics starting from torus creation for increasing coupling strength. Not far from the TR bifurcation the torus is smooth and demonstrates anti-phase type of oscillations as seen in Fig. 3a.

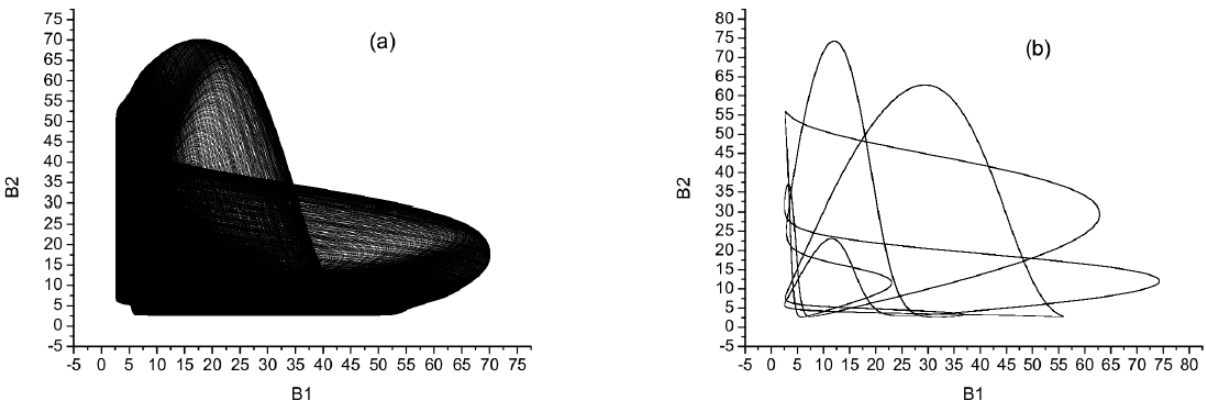

**Fig. 3.** Phase portraits of coexisting torus (a) and LC5:5 (b), **n=3, α=400, Q=0.64**

The boundary of the torus stability in the (**Q-α**)-plane has been calculated by direct integration and/or by the calculations of Lyapunov exponents which clearly capture the torus death. The fate of the system (1) after torus destruction depends on the value of **α** and is presented below.

The unexpected event in the system dynamics is the appearance of the symmetric limit cycle with 5 winding (rotation) numbers, LC5:5 (Fig. 3b), which is not a classical resonance on the torus. It coexists with the torus, and its dynamical behavior as a function of **Q** is controlled by the value of **α**. Figure 4 shows the family of **Q**-continuation plots of the LC5:5 for different values of **α**. Also shown is the resonant LC5:5-R which appears as the low-**Q** pieces for **α** = 400 and 800. For **α** = 400 the LC5:5-R exists for only a very narrow **Q**-range, appearing as a small spot in Fig. 4. Here we focus our interest on the LC5:5 and its development as **α** increases.

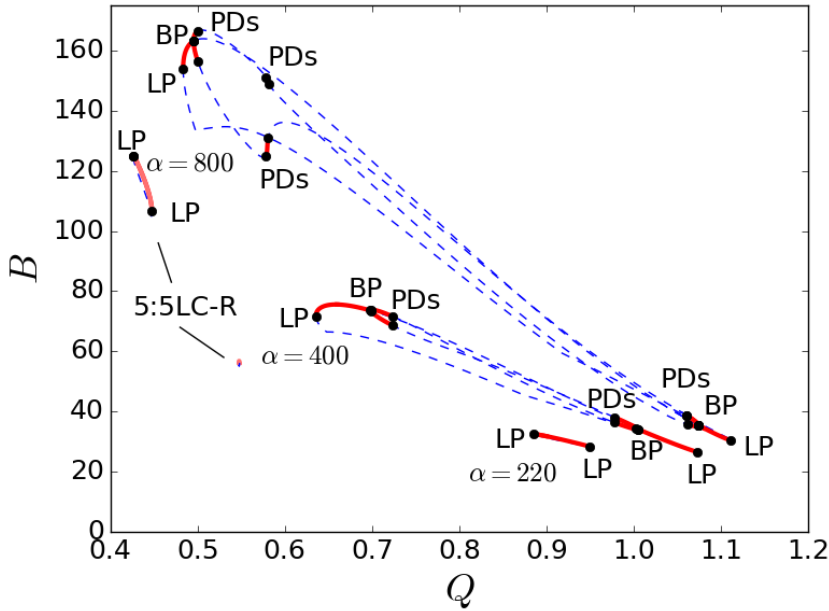

**Fig. 4.** The amplitude of oscillations for LC5:5 as a function of **Q** for **α** =220, 400, 800. Also shown are resonant LC5:5-R occurring for **α** =400 (near the start of Arnold's tongue, see Fig.5) and 800 (two parameter continuations of both LPs, see in Fig.5). BP is the branch point (pitchfork bifurcation) leading to the loss of symmetry of LC5:5; PDs mark the start of period doubling cascades of asymmetric LC5:5.

For the small **α=220**, the LC5:5 is continuously stable and coexists with the torus over the extent of the LC5:5. Increasing **α** dramatically changes the dynamics of the LC5:5, which, first, loses symmetry in pitchfork bifurcation (BP in Fig.4) and, second, goes to chaos via period doubling cascade as seen for α=400, 800 in Fig. 4. At the large Q>1 the system returns to stable APLC using the reverse sequence of bifurcations. At the large **α=800**, the appearance of additional PD bifurcations within the chaos produces a narrow region of stable LC5:5, which is not studied in this work. To reveal the dynamics in detail and over broad intervals of key parameters, 2-parametric (**α, Q**)-continuations were calculated for the boundaries of torus stability, the bifurcations (LP, BP, PD) of the basic regime LC5:5, and other selected influential limits cycles. Figure 5 shows the resulting map of regimes.

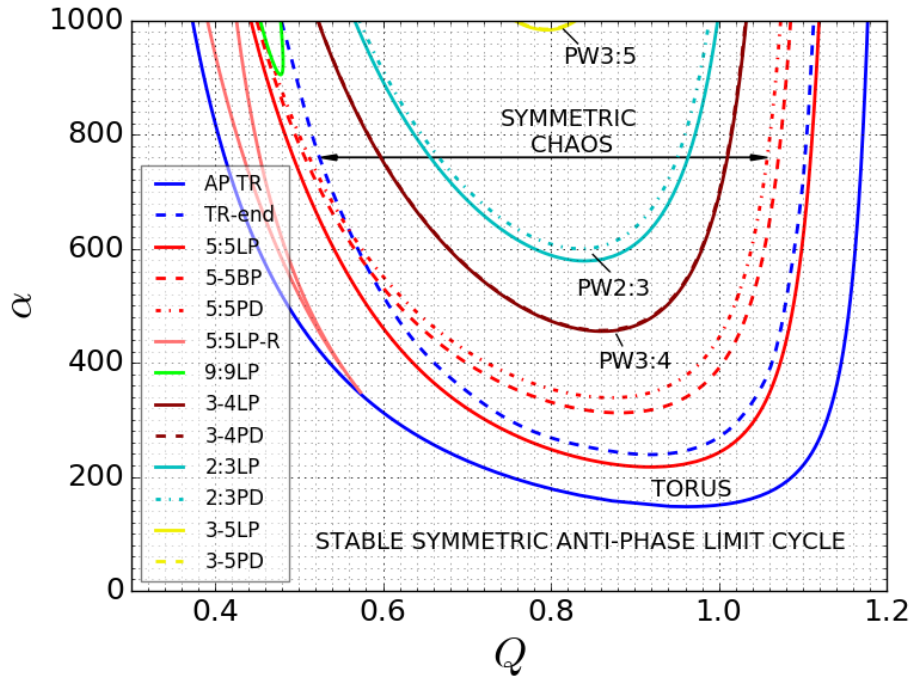

**Fig. 5.** The phase diagram demonstrating the boundaries of stability for basic regimes. For 3:4LC and 3:5LC which are located in intrachaotic periodic windows (PW), separation between LP and PD lines is not apparent at this resolution.

The order of bifurcations looks like a "matryoshka" in the (**Q-α**)-plane, with nested torus, LC5:5, LC3:4, LC2:3, and LC3:5, with the oddity of the strip of overlap due to coexistence of torus and LC5:5. Also shown is the resonance LC5:5-R, a classical Arnold tongue, which is represented here to distinguish the positions and the roles of the two L5:5s. Chaos appears on the (**Q-α**)-map for α > 350 due to the PD cascades of the nonresonant LC5:5. The main impression from this map of regimes for our system of two identical coupled oscillators is the vast area between these PD cascades occupied by chaos. Consider the routes to this chaos.

For 350 < α < 635 (according to Fig. 5), torus destruction takes place before period doubling of the LC5:5. After its destruction, the system goes into the LC5:5, the single attractor creating chaos during further **Q** growth (see Fig. 6 for **α**=600).

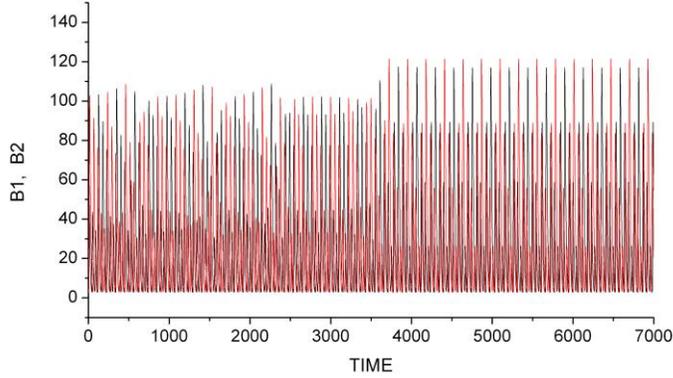

Fig.6. Time series B1, B2 of the torus destruction and the system transition to slightly asymmetric LC5:5, n=3, **α** =600, **Q**=0.5685.

The LC5:5 loses symmetry via pitchfork bifurcation, then undergoes a PD-cascade to chaos. For **α** = 600, the torus "converts" directly to the As-LC5:5 because the end of torus stability occurs at slightly higher **Q** than does the pitchfork bifurcation (BP in Figs. 4, 5) which creates the As-LC5:5 from the symmetric one. (For smaller **α**, the torus stability ends before the BP, meaning the torus "converts" to the symmetric LC5:5). The evolution of chaos is presented as the series of sequential period maps with increasing complexity (Fig. 7):

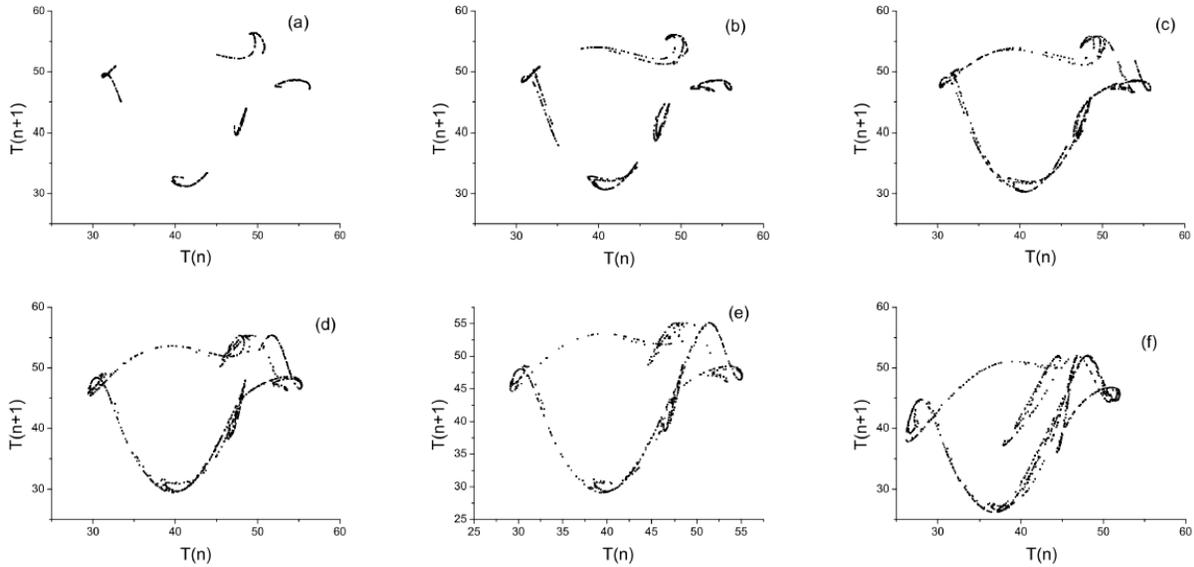

**Fig. 7.** The evolution of sequential period maps **T(n+1) vs T(n)** after the period doublings of AsLC5:5 up to the mature chaos for **α**=600. The value **Q**=0.58 is the end of LC5:5 PD cascade, (a) Q=0.592, (b) Q=0.604 (c) Q=0.614, (d) Q=0.634, (e) Q=0.654, (f) Q=0.744. T(n) are the set of Poincaré return times calculated as the time intervals between the intersections of ascending trajectories of $B_1(t)$, $B_2(t)$ with the line B = 7.

For **α** > 650, the end of torus stability is beyond the PD-cascade thereby allowing both the torus destruction and the period doubling of LC5:5 to serve as routes to chaos. Although the Q-interval of these attractors' coexistence is small, it is interesting to trace the development of chaos in the presence of the torus covered by resonant cycles. For example, Figure 8 shows the behavior of two sets of Lyapunov exponents for **α** = 800: black, blue, pink solid lines correspond to the calculations started from LC5:5 while red, green, navy dashed lines demonstrate LEs of regime started from the torus.

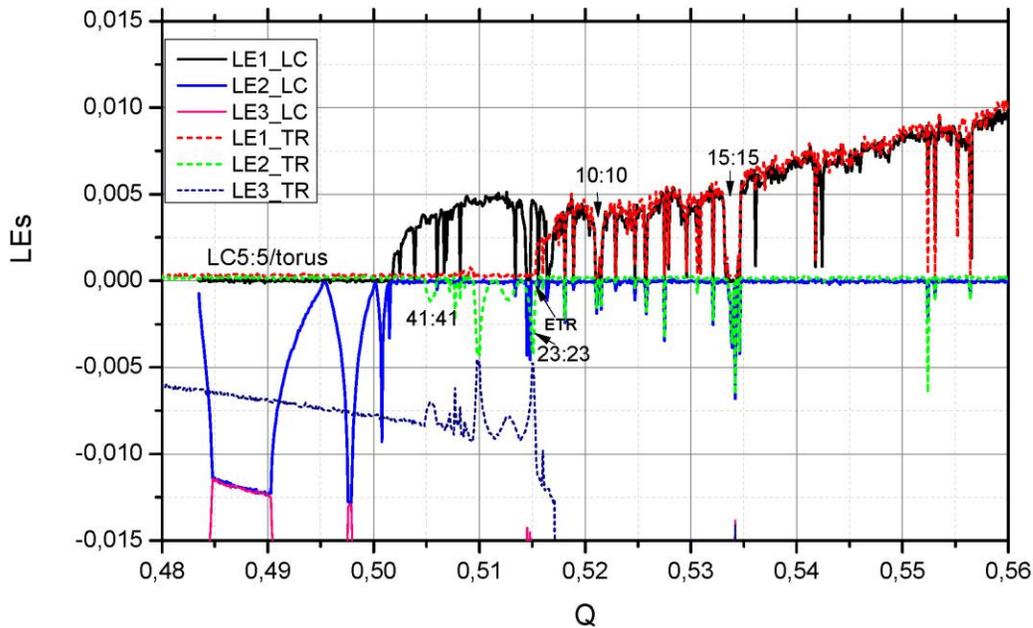

**Fig.8.** Two sets of three LEs as a function of **Q**: start from 5:5LC (solid black, blue, pink lines), start from torus (dashed red, green, navy lines). Some winding numbers for resonant cycles on torus and the cycles in periodic windows in chaos are presented. ETR marks the end of torus. **α** = 800, **Q**-step=0.0001. Here and in other Figures for LEs for two regimes the values for one attractor were artificially shifted upward by 0.0001 or 0.0002 to distinguish lines near LE=0.

The plots of the two main LEs (solid black and dashed red lines in Fig. 8) show that the LC5:5 and the torus are stable and coexist up to the chaotization of the LC5:5 near **Q**=0.5015 (positive bump of black curve). This weak chaos we denote as 5:5-chaos. The torus remains stable as **Q** increases, however, the **Q**-region close to the torus destruction (marked by ETR in Fig. 8) contains many resonant cycles, which manifest themselves as several dips of the dashed green curve (LE2-TR). The winding numbers of two resonant cycles are marked in Fig.8, with the LC23:23 being the starting point of torus destruction. Thus, there is interesting coexistence of the torus and the 5:5-chaos.

To illustrate the torus transition to chaos we investigated sequential period maps T(n+1) vs. T(n) in the region of the torus destruction. The first map for **Q** before transition (Fig. 9a) is not smooth but it is, nevertheless, a closed curve while the second map (Fig. 9b) contains additional points. Their positions in Fig. 9b indicate the start of torus destruction, after which the evolution of chaos is very similar to that presented above for **α**=600 (Fig.7**b**).

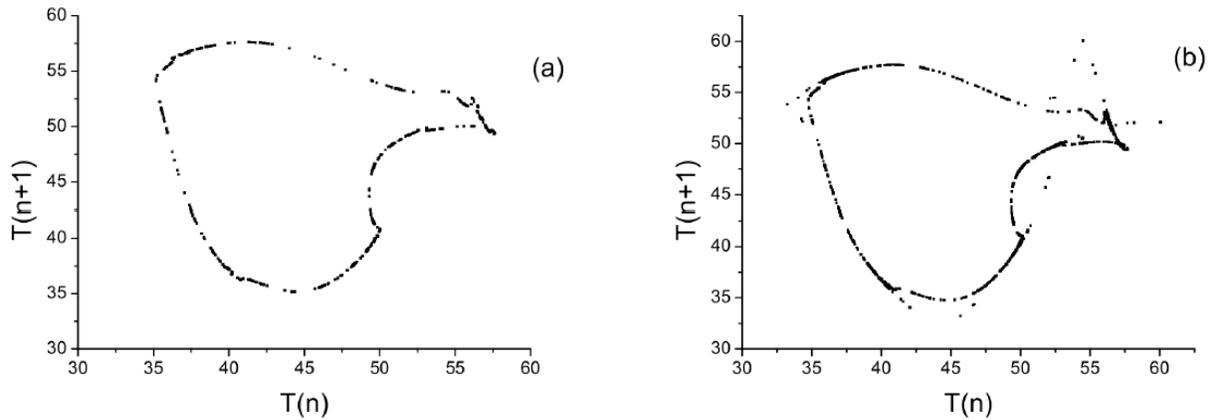

**Fig. 9.** T(n+1) vs. (Tn) maps: (a)- for torus at **Q**=0.5155 just after LC23:23; (b)- for transient trajectory between torus and chaos **Q**=0.5172. **α** = 800.

The coexistence of chaos and a non-smooth torus bearing many resonance cycles seems to be a non-trivial dynamical phenomenon. The **Q**-interval between the chaotization of LC5:5 and the end of the stable torus becomes longer if parameter **α** increases (see lines 5:5PD and TR-end in Fig. 5) thereby opening other routes to chaos.

Increasing **α** to 900, we find that the 5:5-chaos is no longer stable over the entire interval to the ETR. The coexistence ends at the death of the 5:5-chaos, where the system transitions to the torus. Figure 10 shows the time series of the transition from weak 5:5-chaos to torus, and its sequential period map. The map contains points for both regimes: the closed curve with small loop is the map of torus while the combination of the five separate elements is the map of the 5:5-chaos just before the loss of its stability. The torus is then stable over a small Q-interval until the appearance of a resonant LC23:23 (**Q**=0.4925) followed by the system's chaotization similar to that presented in Fig. 9b for **α**=800.

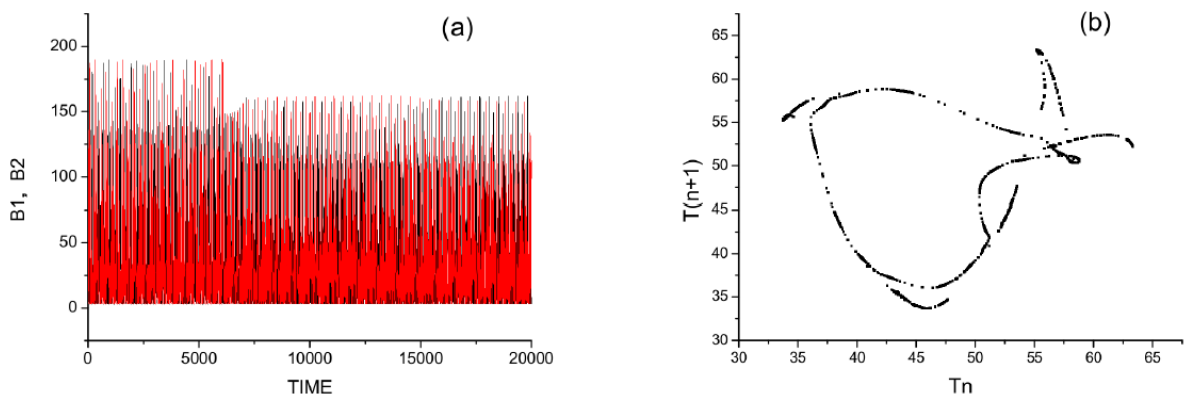

**Fig. 10.** The system transition from 5:5-chaos to torus for **α**=900, **Q**=0.4808: (a)-time series; (b)- sequential period map T(n+1) vs (Tn), closed curve – torus map; five separate pig-tails – map of the weak 5:5-chaos.

Increasing **α** beyond 900 causes the evolution described above for α=900 to be interrupted by the appearance of the relatively broad (**Q-α**)-island created by the asymmetrical cycle As9:9LC seen on the top left in Fig. 5 (green line) and shown zoomed-in in Fig. 11.

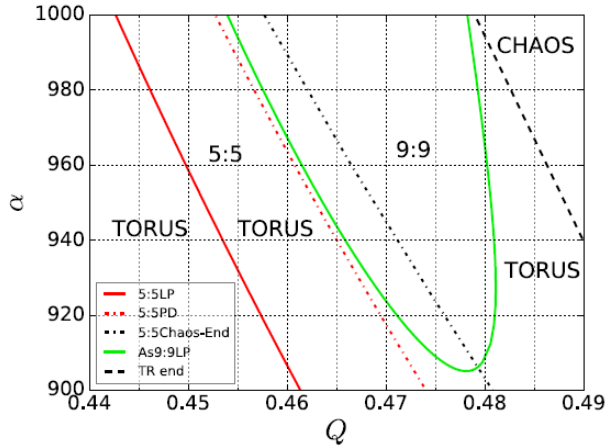

**Fig. 11.** The zoom of the (**Q-α**)-map where the route to chaos via AsLC9:9 takes place.

As an overview of the effect of the AsLC9:9 island we consider **α**=1000. Figure 11 indicates that as **Q** increases from the 5:5PD, the LC5:5 undergoes period-doubling to chaos which coexists with the As9:9LC until the chaos becomes unstable and the system then transitions to As9:9LC at **Q** about 0.458. Prior to the transition, the degree of chaos formation may be estimated from the sequential period map (Fig. 12a) while the system transition from the chaos to the As-LC9:9 is visualized in Fig. 12b.

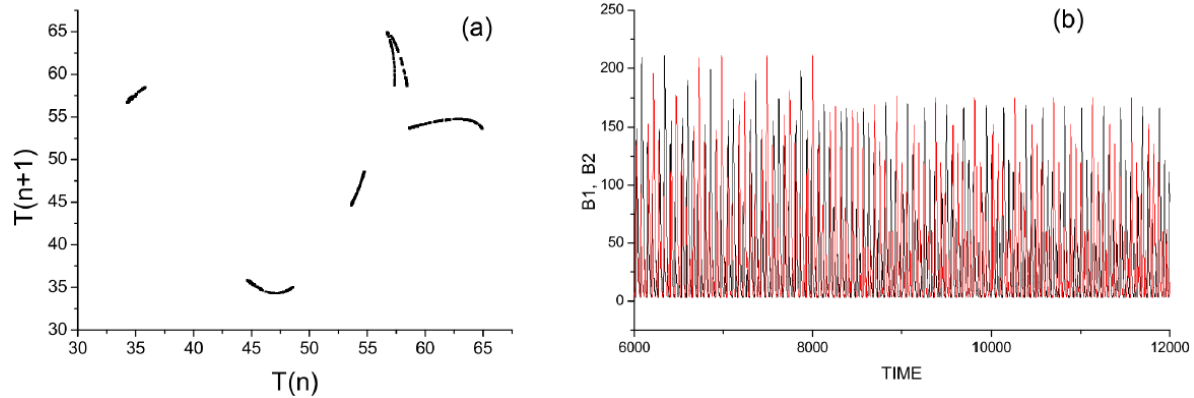

**Fig. 12.** (a) T(n+1)vs T(n) for Q=0.4577 just before the weak chaos transition to asymmetric cycle LC9:9; (b) the time series of this transition after very small increase of **Q**. **α**=1000.

The Lyapunov exponents in Fig. 13 for trajectories started from the LC5:5 and from the torus show the short Q-interval of coexistence of the weak chaos originated from LC5:5 and the As-9:9LC, followed by the death of the chaos at **Q**=0.458 leaving the As-9:9LC as the sole stable dynamical behavior. The As9:9LC then loses stability near **Q**=0.478 generating chaos which gradually develops into the mature chaos with sequential period map like that presented in the Fig. 7.

Comparing the LE graphs for **α**=800 and 1000, Figs. 8 and 13, it is seen that the As9:9LC at **α**=1000 plays the same role as the torus at **α**=800 regarding coexistence with 5:5-chaos, and being the sole attractor during final evolution to chaotization. The similar roles are expected since the As9:9LC is an unusually broad island inside the torus.

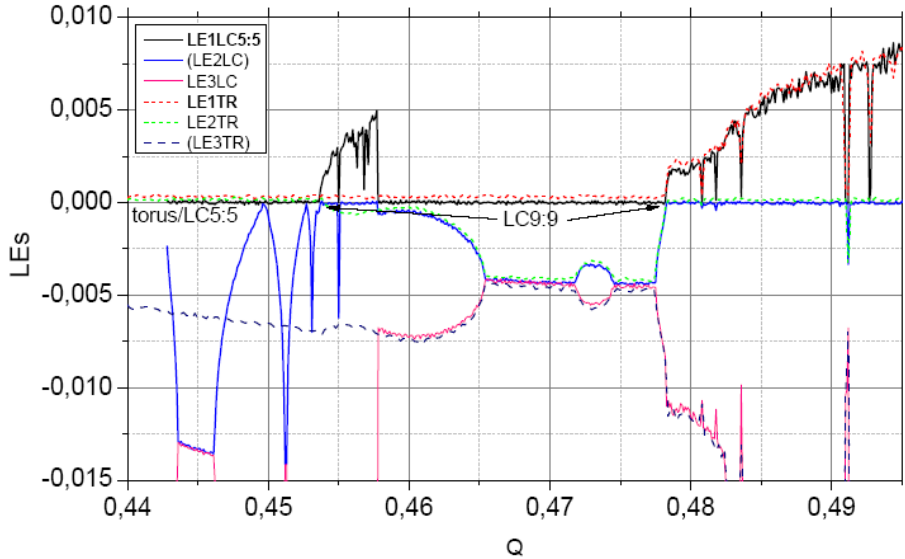

**Fig. 13.** Two sets of LEs for trajectories, started from LC5:5 (solid black, blue, pink lines) and from torus (dashed red, green, navy lines), as a function of **Q**, **α**=1000.

To investigate more thoroughly the route to chaos after crossing the regime of stable AsLC9:9, the LEs for torus continuation for **α**=930 have been calculated with a very small Q-step (0.00001) and are shown in Fig. 14.

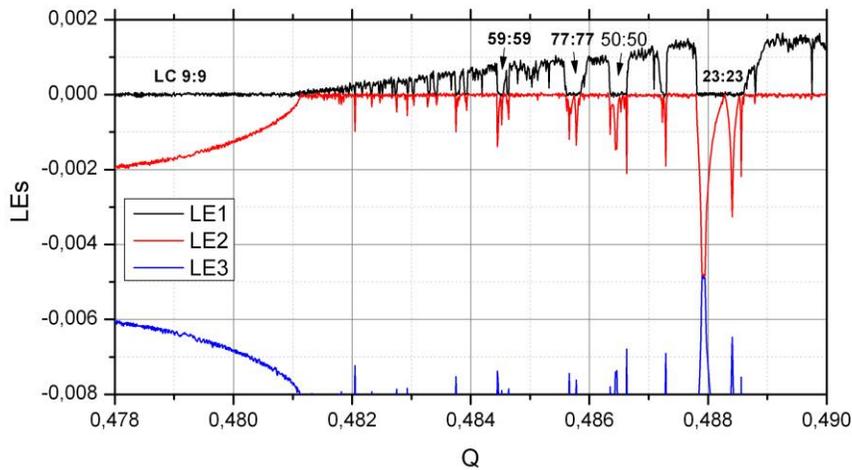

**Fig.14.** The detailed evolution (**Q**-step is 0.00001) of LEs starting from AsLC9:9 up to the appearance of irreversible chaos formation after the period doubling of LC23:23, **α**=930.

Behavior of the main LEs shows that the AsLC9:9 is stable up to its LP (**Q**=0.4811) after which two LEs are zero demonstrating the system transition to the ergodic torus. However, the torus is not clearly exhibited because it is covered by tightly packed resonant cycles whose beginnings of chaotization are unsuccessful until the period doubling of the LC23:23 which leads to creation of the robust chaos after Q=0.492 where the first LE amplitude surpasses 0.002. This scenario of the torus with tightly packed resonant cycles leading to chaotization at the ETR occurs for **α** up to about 1000. It is different from the LC5:5 period doubling chaotization which dominates for **α** < 800. However, after evolution of the chaos with increasing **Q**, both ways produce similar chaotic regimes according to the values of LEs and the structure of recurrence time maps.

After chaotization the appearance of many periodic windows is observed over the vast (**Q-α**)-area. In most of them limit cycles are symmetric with identical winding number, LC*n:n* and identical amplitudes of oscillations in the two oscillators. However, the dynamics in some windows deserve special attention because they contain spatially asymmetric limit cycles which are not typical for identical oscillators. The boundaries of AsLC3:4, AsLC2:3, and AsLC3:5 are pictured in Fig. 5 while the periodic windows with other partial asymmetric LCs are too narrow to be found with the chosen numerical accuracy used in Fig. 5. Although the bands with stable AsLCs are narrow (see, for example, interval between 2:3LP and 2:3PD in Fig. 5) their unstable branches cover large areas of the (**Q-α**)-plane forming asymmetrical elements in the chaotic time series. A typical example of the chaotic trajectories with fluctuations of asymmetric "polarity" is presented in Fig. 15(a) as well as a zoom in Fig. 15(b) showing the long lived unstable orbit LC2:3. It is notable that for two coupled identical oscillators, the version of QS coupling used in system (1) has disrupted the homogeneity of oscillations inside chaos and has generated very different values of recurrence times.

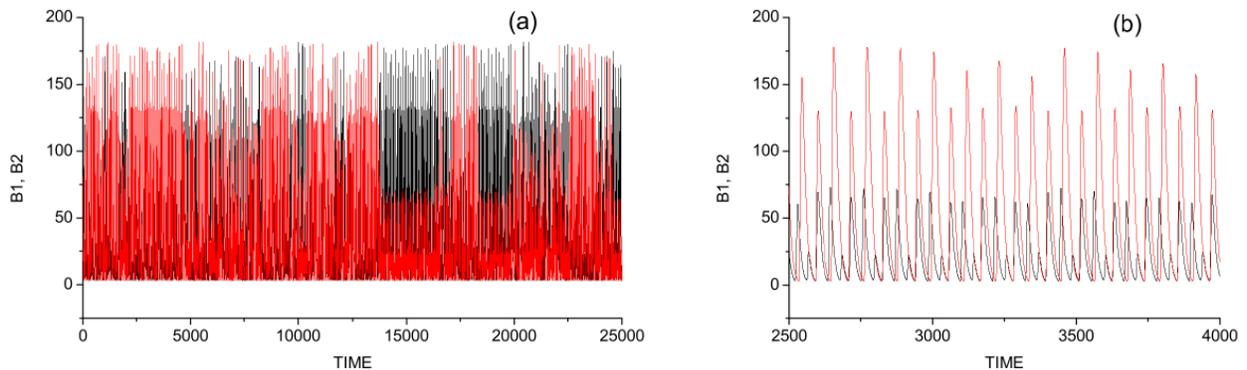

**Fig. 15.** Time series of symmetric chaotic regime for n=3, **α** =1000, **Q**=0.58 (a) and a zoom of "polarized" part from this time series (b) showing the long-lived unstable 2:3LC.

Thus, for **α** < 1000 two identical Repressilators indirectly coupled via diffusion of autoinducer demonstrate several important dynamical features: (i) the existence of a very large area of parameter-space where the appearance of self-organized quasiperiodicity occurs, which is unusual for coupled *identical* oscillators; (ii) the coexistence of torus and non-resonant LC5:5, each of which initiate the development of chaos in different large areas of the parameter plane; (iii) the existence of periodic windows with asymmetric limit cycles, the unstable branches of which cover large regions of the parameter space, ensuring the "polarization" of chaotic time series.

Further growth of **α** (next section) stimulates the appearance of the other asymmetric limit cycle 1:2, which, in turn, produces the coexistence of regular, quasiperiodic and chaotic, symmetric and asymmetric regimes over very large parts of (**Q-α**)-plane.

## 3.2 High-strength Oscillators: α > 1000

In the previous section, we saw that the main LC5:5 and the torus provide routes to chaos for most of the region for **α**<1000. For 800<**α**<1000 the weak chaos that arises from the period doubling of the LC5:5 is stable for only a short Q-interval (see Fig. 13). When this chaos becomes unstable the system transitions to either the torus or As9:9LC depending on the location in the (**Q-α**)-map. However, the As9:9LC becomes unstable with further increase of Q and the system transitions to the torus which then provides the final stage of the route to chaos emerging at ETR.

Figure 16 shows the region of the (**Q-α**)-map in which the LC5:5-Res and the As9:9LC island are the important dynamics interacting with the torus to provide routes to chaos as **α** increases beyond 1000. Surprisingly, the external boundary of AsLC9:9 has a smooth closed form and is limited to α<3200, while the internal marking of stability areas is delineated by additional LP and PD lines.

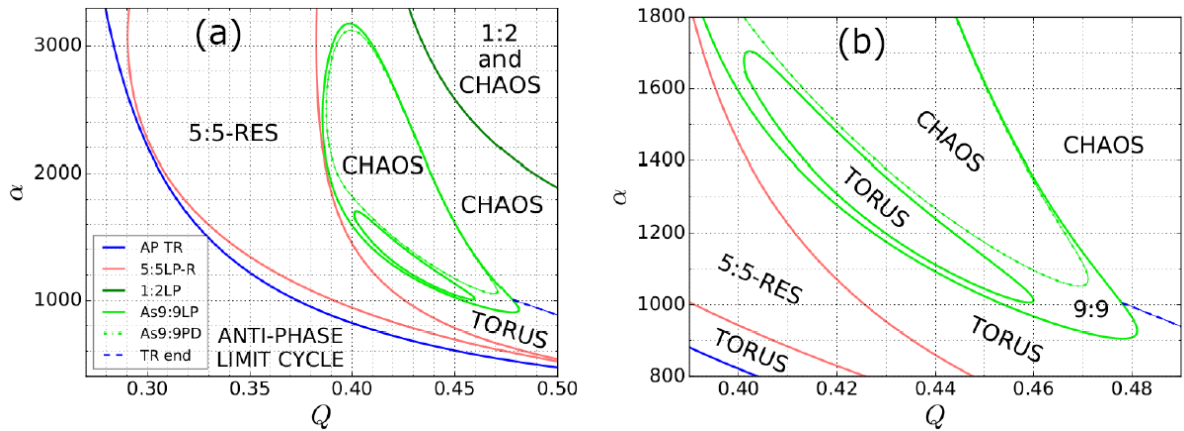

**Fig. 16.** (**Q-α**)-map of the AsLC9:9 islands, 1:2LP- the boundary of LC1:2 stability, and the resonance LC5:5-R: (a) boundaries of basic regimes and (b) zoom showing the structure of torus and AsLC9:9 locations (compare with Fig.17).

The route to chaos for **α**=1200 is shown by the LEs graph in Fig. 17, in agreement with the boundaries encountered along the α = 1200 line in Fig. 16b. The LC5:5-Res loses stability at its LP and the system transitions to the torus as indicated by a second LE with zero amplitude. The torus has the set of resonant limit cycles (e. g. 14:14, 23:23) and the asymmetric LC9:9 at the negative dips of the second (green) LE. In contrast to the case in the previous section for **α** =1000 where LC9:9 is continuously stable, for α = 1200 there are 3 narrow stable LC9:9 regions separated by torus and by chaos (main LE > 0). Figure 16b shows the stable torus region delineated by interior 9:9LP structure separating the first two stable LC9:9 regions (see the small oval in Fig. 16b), and the period-doubled chaos region separating the next two. The chaos emerging from the period-doubling cascade of the LC9:9 matures, as indicated by the growing main LE. The growth is interrupted by the narrow periodic window which contains the period-halving cascade back to stable 9:9LC in the Q-span just prior to the high-Q LP.

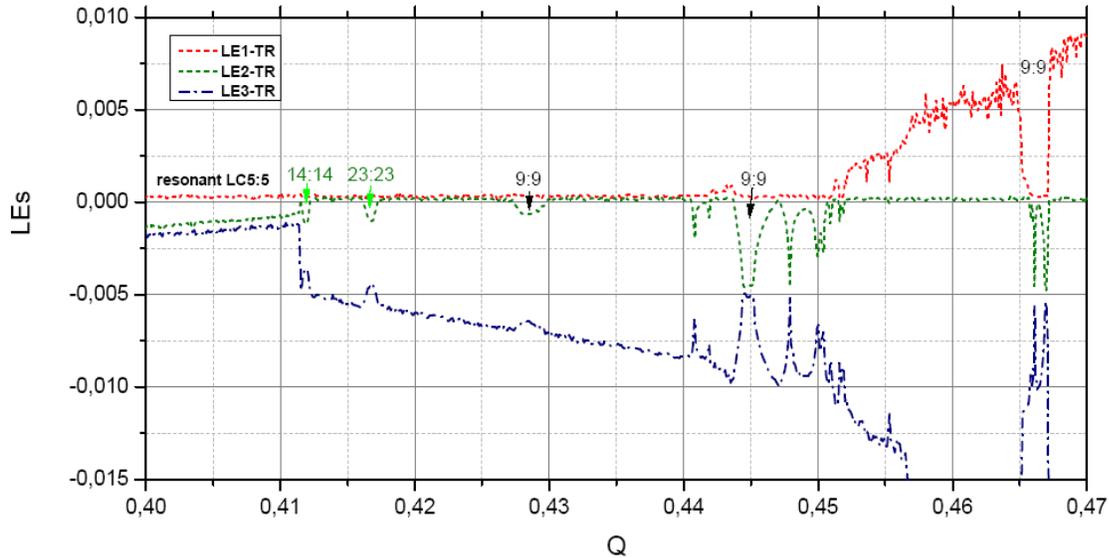

**Fig. 17.** LEs as a function of **Q** (step=0.0001) for trajectory started from resonant LC5:5-Res. $\alpha=1200$.

As $\alpha$ increases up to 2200, the Q-interval with stable AsLC9:9 shortens (Fig. 16) and the chaos emerged from period-doubling starts at smaller coupling strengths. This symmetric chaos extends to larger coupling strengths where it encounters and coexists over a broad Q-span with a new asymmetric limit cycle, AsLC1:2, shown in Fig. 16 with dark green line. The time series of both regimes are presented in Fig. 18 which demonstrates the sharp transition from symmetric chaos to the AsLC1:2 which occurs at the end of the coexistence region.

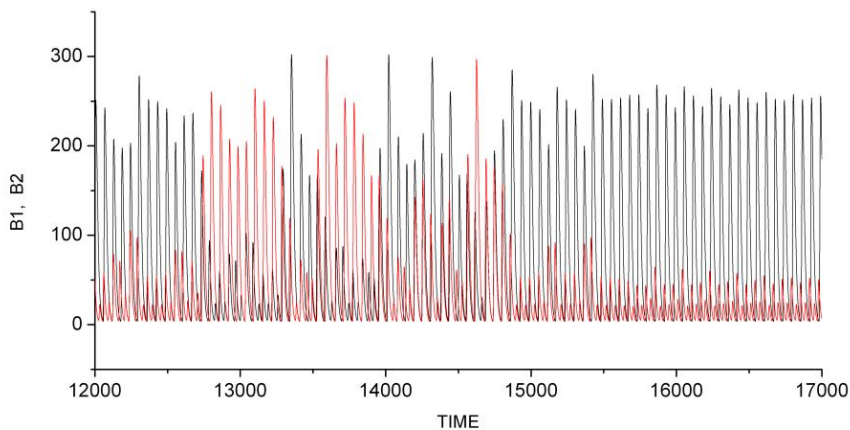

**Fig. 18.** The time series of transition from chaotic regime to stable LC1:2 at Q=0.568, for $\alpha$=2000. The small fluctuations of amplitudes after the transition are the transient.

By analogy with Fig. 4 for the LC5:5, we calculated for different $\alpha$, the family of LC1:2 Q-continuations shown in Fig. 19 which delineates the areas of stability and reveals the basic bifurcations of this new limit cycle.

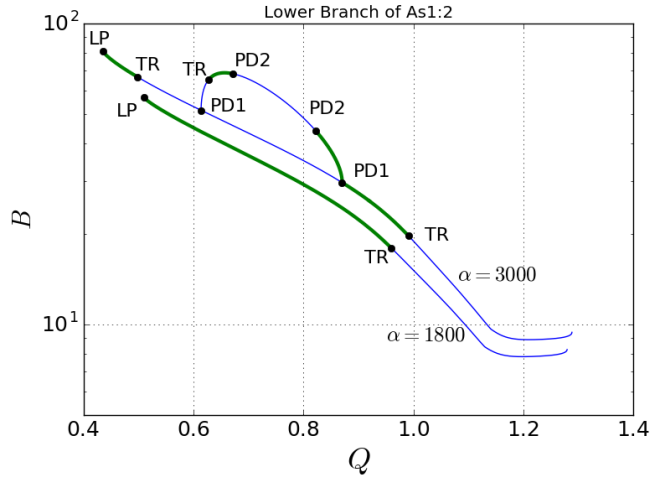

**Fig. 19.** The lower branch of one-parameter Q-continuations of LC1:2 amplitude for $\alpha$ =1800 (LC1:2 is stable from LP to TR) and for $\alpha$=3000 with TR bifurcations and PD cascades.

Throughout a large region of coupling strengths the external boundaries of 1:2LC stability are determined by a limit point at the low-Q end and torus bifurcation at the high-Q end, while the internal structure of the phase diagram is mainly controlled by the locations of torus and period doubling bifurcations.

Two-parameter continuations of bifurcation boundaries of important limit cycles, including the LC1:2 in Fig. 19, are presented in Fig. 20, which is an extension of Fig. 5 for large $\alpha$. The boundary of stability of the symmetrical chaos, which coexists with LC1:2, is calculated by direct integration of the time series or LEs and is presented in Fig. 20 as a black dashed line.

It is noteworthy that over a wide range of $\alpha$, from 1200 to 2500, chaos is not restored in the central range of Q-values (0.65 to 0.95). Instead, after the end of the chaos emerged from PD of AsLC9:9, there is a broad area dominated by stable LC1:2 and LC2:4. It is only when $\alpha$ surpasses 2500 when the PD-cascade of the LC1:2 creates chaos.

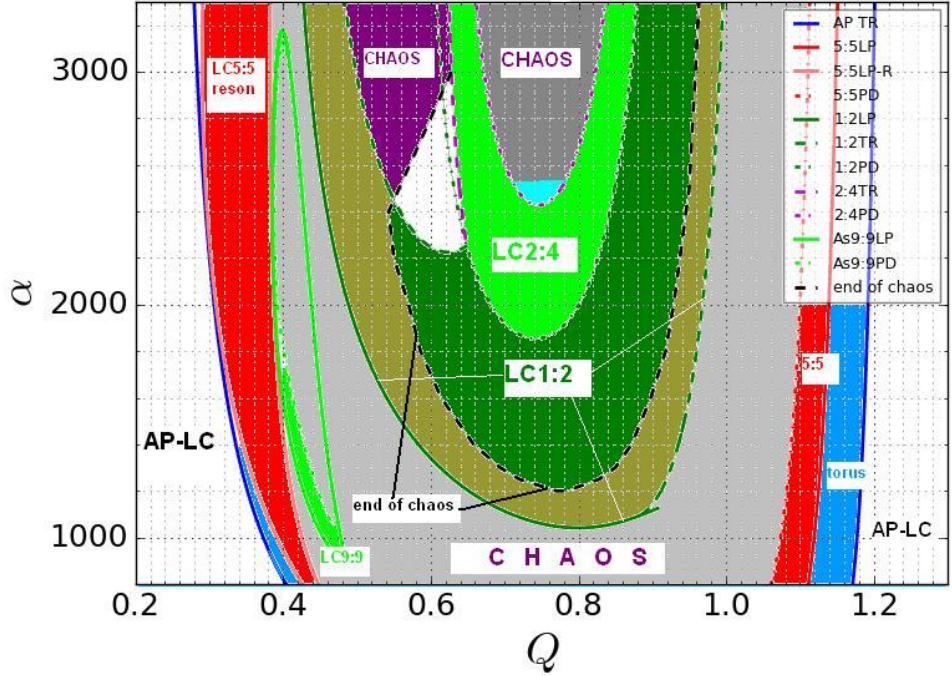

**Fig. 20.** Q-α regime map for α > 1000. Only those boundaries relevant to chaos evolution are shown. The white area adjacent to LC1:2 and LC2:4 contains many asymmetric limit cycles, see Fig. 22 as example.

For α > 2500 and the intermediate values of coupling strength (0.65<Q< 0.85), the system (1) demonstrates the classic route to chaos via period doubling cascades of the AsLC1:2. The chaotization after period doublings results in the restoration of spatial symmetry of the attractor as a whole because the amplitudes of the Repressilator's oscillations become the same.

However, this formal symmetrization masks the real temporal asymmetry of the chaotic trajectories. The typical trajectory presented in Fig. 21 clearly demonstrates the stochastic long-lasting intervals with the fixed polarity.

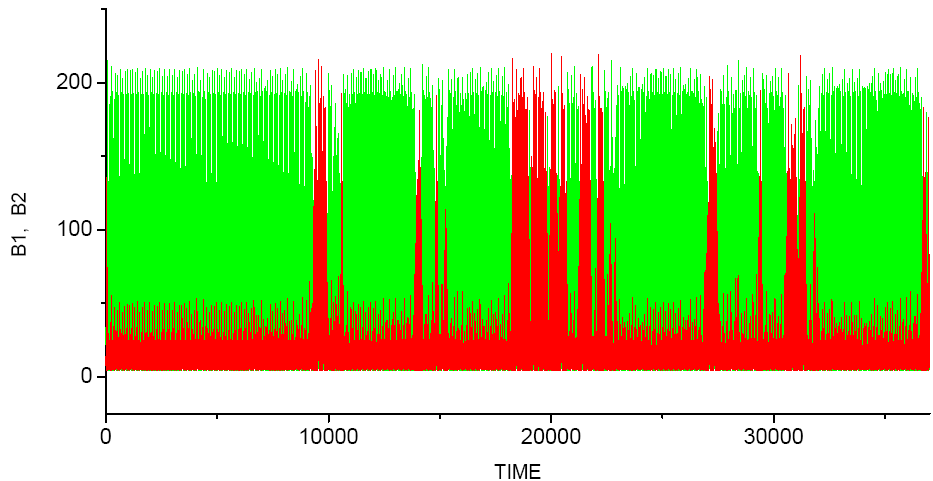

**Fig. 21.** The random example of the time series demonstrating the chaotic switching of polarity, *α*=3000, **Q**=0.797.

The existence of long almost deterministic asymmetric orbits within a chaotic time series can change the widely-held view of chaos as a short-correlated irregular process. Because the asymmetric LC is the starting attractor for the development of chaos, there are large parameter regions in which the lengths of time series used for LE calculations should be chosen taking into account the characteristic durations of periodic orbits embedded in chaos. The LEs presented below were calculated using very long samples to ensure reliable averaging.

Further increase of **Q** restores the asymmetry of dynamical regimes due to the reverse cascade of PD bifurcations which takes the system back to LC1:2. The AsLC1:2 is stable up to the high-**Q** 1:2TR seen in Fig. 20, where it then loses stability and the symmetric chaos is again the single attractor of our system. An interesting qualitative phenomenon of the system is the continuous strip of coexistence of the symmetric chaos and the LC1:2 formed by the end-of-chaos boundary inside the 1:2TR for high-**Q** and inside the 1:2LP for low-**Q**. The coexistence of symmetrical chaos with asymmetrical limit cycles is a remarkable peculiarity of the system (1). The manifestation of this coexistence for the low-Q region is different from that described above for the high-**Q** region. At the low-**Q** region bifurcation analysis discovered one more special chaotic area bounded by the torus bifurcations of LC1:2 (see lines low-**Q** 1:2TR and 2:4TR in Fig. 20).

Consider the dynamics for the case *α* =2500**.** For Q-values beyond the LP1:2-line the system (1) can be found in three coexisting states: LC1:2, LC3:6 (see below) and the symmetric chaos. The evolution of the LC1:2 is presented in Fig. 22 where two LEs=0 starting at Q=0.542 indicates the start of the torus, in agreement with the 1:2TR line in Fig. 20. The resulting torus is asymmetric, and its development towards chaos is very limited for *α*=2500. Instead, the asymmetric branch contains a sequence of asymmetric limit cycles: 5:10, 7:14, 9:18, 12:24, 13:26 and the system demonstrates the transition to LC2:4 via 2:4TR bifurcation (Q=0.64) as prescribed by the map in Fig. 20.

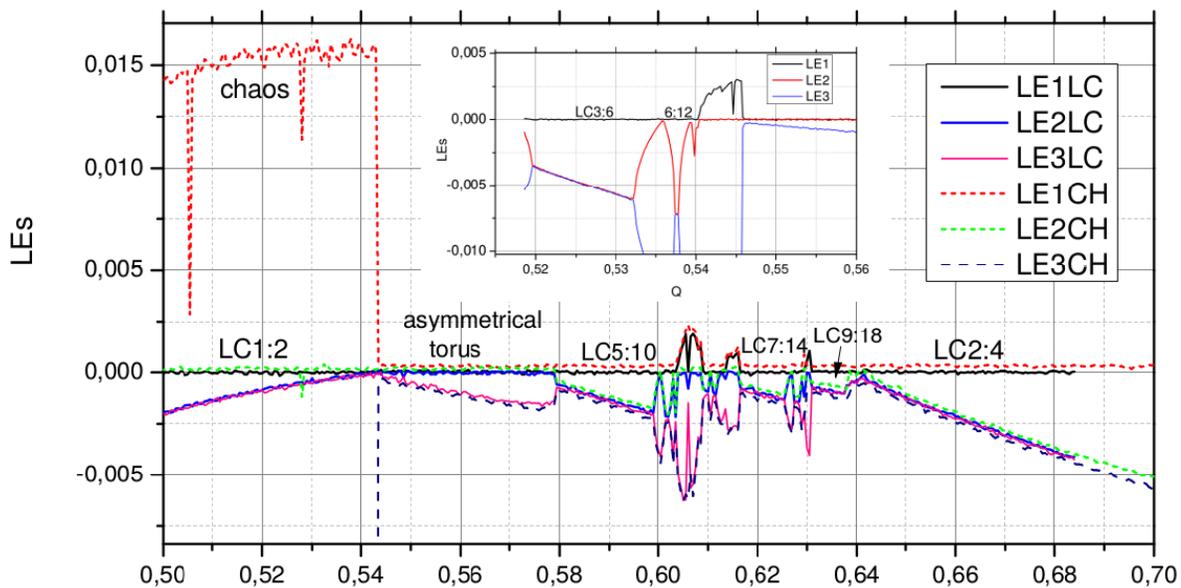

**Fig.22.** The dependence of LEs on **Q** (step=0.0005) for n=3, **α** =2500. The starting regimes are: LC1:2- solid lines and the symmetric chaos – dashed lines. The insert is the LEs vs Q for regime LC3:6, which occurs at Q=0.518 and overlaps with LC1:2 and chaos.

As for the symmetric chaos, for α=2500 it loses stability near the same Q-value where the LC1:2 lost stability, 1:2TR (Q=0.542), resulting in the system switching to an asymmetric weakly chaotic regime originated from the period doubling cascade of the LC3:6 (insert in Fig.22). This sudden loss of chaotic symmetry is illustrated in the time series in Fig. 23. Then the system evolves up to the transition of the weak asymmetric chaos to asymmetric torus at Q=0.5457.

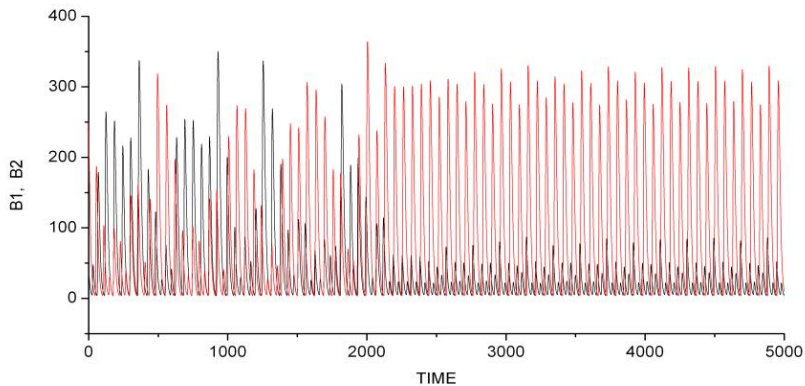

**Fig. 23.** The symmetric chaos loses stability and the system switches to asymmetrical weak chaos originated from PD-cascade of LC3:6. **α**=2500, **Q**=0.5428

For larger values of **α,** the system demonstrates a wider area of stability for the symmetrical chaos. For example, at **α**=3000 the symmetric chaos is stable up to the transition to LC2:4 (see below the details of this transition) and it shares the phase space with LC1:2 and its asymmetric derivatives. The evolutions of LEs for chaos and LC1:2 are presented in Fig. 24 where the new regime LC3:6 is observed.

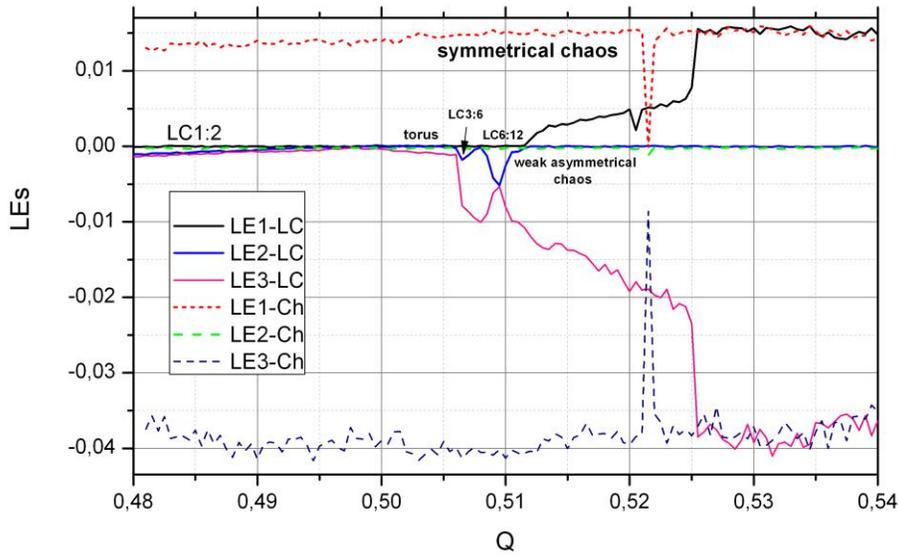

**Fig. 24.** Two sets of LEs vs Q for both regimes: start from LC1:2 – solid lines, start from chaos – dashed lines.

The LC3:6 is another exotic example of the coexistence of the symmetric chaos with an asymmetric limit cycle (see Fig. 24). LC3:6 is the tripled version of LC1:2 (see Fig. 25a) and is stable in a narrow band along the 3:6LP line (Fig. 20). It undergoes period-doubling bifurcation to chaos, first to a weak asymmetric chaos followed by the transition to strong symmetric chaos as directly observed in Fig. 25(b) and supported by the sharp increase of the first LE (black line in Fig. 24 near **Q**=0.525). The LC3:6 is not a periodic window within the symmetric chaos. Instead, it coexists with the symmetric chaos and with the LC1:2.

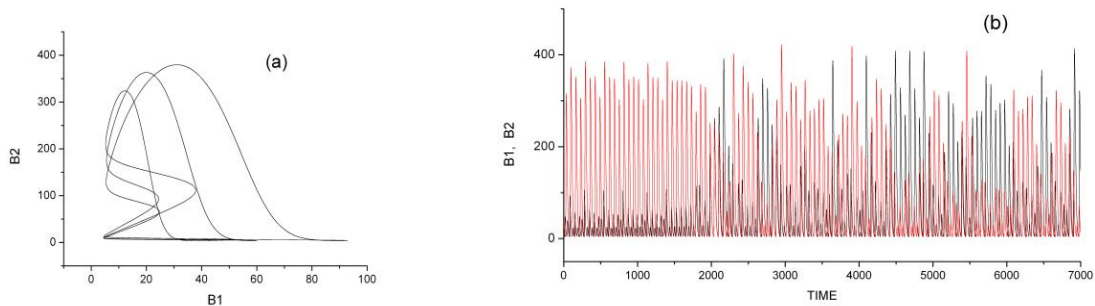

*Fig. 25*. (a)- the phase portrait of LC3:6, **Q**=0.507;(b)- the transition of weak asymmetric chaos to strong symmetric one, **Q**=0.5249. **α**=3000.

Figure 24 shows that the chaos from the period-doubled LC3:6 matures and merges with the symmetric chaos. The further evolution of this single chaotic attractor is shown in Fig. 26. The sym-chaos loses stability, making a sharp transition to LC2:4. The period doubling of LC2:4 restores chaos, which is stable over the large interval of coupling strength up to the high-**Q** period-halving cascade. The system sequentially crosses the regions with LC1:2, LC1:2+chaos, pure chaos and torus, eventually reaching the TR-bifurcation back to stable anti-phase LC (see Fig. 20).

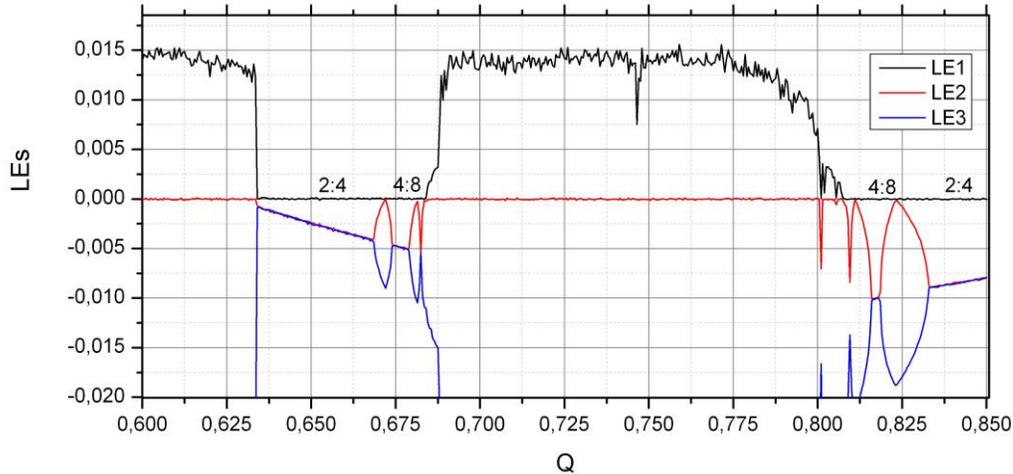

*Fig. 26*. LEs vs Q starting from the symmetric chaos. $\alpha$=3000.

## 4. Discussion

There are a huge number of papers concerning the collective behavior of nonlinear oscillators published since Huygens's study of coupled clocks (see e.g. [1] and references there in). The recent discovery of "chimeras" is an important and impressive example of the role of the coupling type (see review [11]) in the formation of unexpected dynamical regimes. Although the simple version of "quorum sensing" mechanism can be used to connect oscillators of different nature [40, 41], it is most effective for the coupling of genetic oscillators imbedded in cognate bacterial medium. The genetic networks are constructed from many nonlinear elements, each of which may be engineered to accept the inter-oscillator signal produced in accordance with the activity of other genes. The cooperation of this local property of genetic networks with the indirect global coupling via signal molecule diffusion provides many possible combinations of coupling schemes.

The design of our system, like that of the other versions of the model for coupled Repressilators [27, 30-33], is the simplest one for 3-gene networks with one autoinducer. Being produced under control of gene "b" promoter, the autoinducer activates the expression of downstream gene "c". Despite linear diffusion of the autoinducer, the difference in gene "b" and "c" positions inside the Repressilators' ring provides a completely possible, but a nonstandard coupling scheme. The oscillations of repressors $B_1$ and $B_2$ concentrations are first converted to those of the signal variables $S_1$ and $S_2$ with some time delay. Then, the signal molecules mix according to a quasi-steady state approximation and transfer the impact of the superposition of delayed $B_1$ and $B_2$ oscillations onto the production rates of $C_1$ and $C_2$. As a result, the effective frequencies of oscillations with the superposition, as well as their dependences on the model parameters, are different from the internal frequencies of the Repressilators. As a result, the map in Fig. 27 over the entire interval of **α** focuses on the major unexpected collective modes discovered and their interrelations.

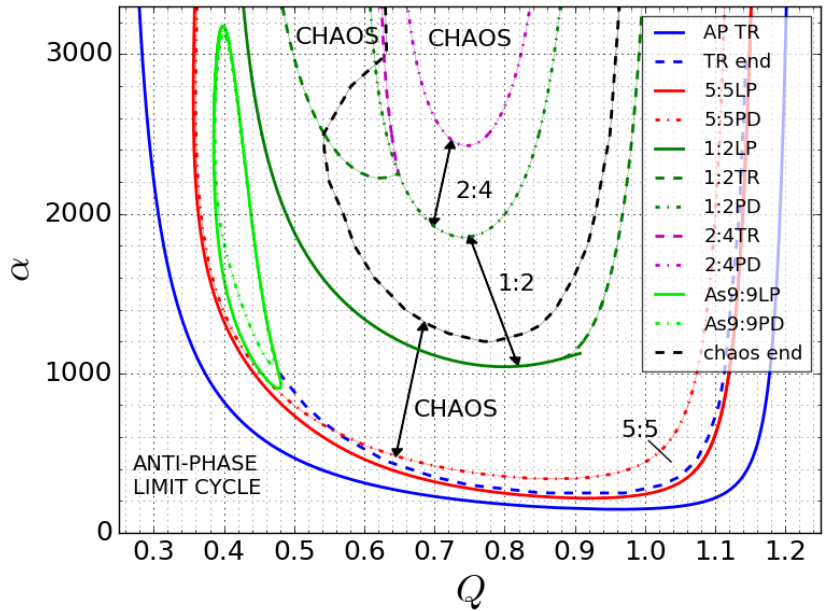

**Fig. 27.** Map of the major unexpected collective modes for the two identical QS-coupled ring oscillators. The paths to chaos are presented in previous Sections.

The external boundary of the large parameter area with the rich set of attractors is formed by the line of torus bifurcation of the anti-phase limit cycle. The appearance of torus in the system of two coupled *identical* oscillators in the absence of an external periodic perturbation is a rare event which requires specific choice of coupling term [42, 43]. The internal boundary of the torus stability (ETR-line) up to **α** = 1000 is presented in the map Fig. 27 and it consists of two parts separated approximately at $\alpha = 800$. The loss of stability after the upper part of the ETR line (800<**α**<1000) is a result of period doubling bifurcations of resonant cycles leading to the chaotization of the system. Along the lower part of the boundary the torus is pushed from the phase space by the LC5:5.

Although there are no indications of its bifurcation origin within the considered areas of parameters, the LC5:5 is of principle importance for the dynamics below **α**=600. It is not a resonant cycle limited by Neimark-Saker bifurcations. In contrast, up to **α**=600 it period-doubles creating a very large area of chaos. For the larger **α**, its role in chaos production is diminished, giving way to torus destruction as a path to chaos. Over the large interval above **α**=1000 a new slightly asymmetrical LC9:9 is located within the closed area (Fig. 27, green line) and the chaos is a result of its period doubling bifurcations. Omitting some details concerning the dynamics of resonant cycles inside the low-Q region adjacent to the TR-line, we concentrate on the chaos creation and conclude that in the system (1) the origin of symmetric chaos is the period doubling bifurcations of different limit cycles: resonant cycles on torus, symmetric LC5:5 and asymmetric

LC9:9. Despite the identical nature of the coupled Represilators, the variability of return times and amplitudes of variables in the 5:5 and As9:9 limit cycles are significant.

Further development of chaos over (**α-Q)-**plane demonstrates the surprising formation of asymmetric limit cycles in narrow periodic windows, e.g. 3:4, 2:3, 3:5, which return to chaos via period doubling bifurcations.  The stability of chaos unexpectedly disappears (dashed black line in Fig. 27) and the system jumps to the asymmetrical LC1:2 which is unlike standard limit cycles within PWs.  The LC1:2 (dark green line in Fig. 27) coexists with chaos and evolves independently generating its own chaos via the period doubling bifurcations ending in the restoration of the symmetrical chaos over a very large area of parameters. Although there are known examples of chaos formation owing to the interaction of identical oscillators [44-47], the indirect QS-coupling gives rise to new uncommon limit cycles being starting points for chaos and its complex evolution over the parameter plane. Classical bifurcations of limit cycles -- period doubling and torus formation-- lead to standard processes of system chaotization. However, under multistability conditions, they generate an unusual phase diagram with domination of chaos and asymmetric attractors.

How general are the set of attractors and the observed structure of the map bearing in mind the unavoidable deviations from perfectly identical oscillators and the reasonable shifts of other model parameters, e.g. Hill coefficient *n* or the AI-induced activation rate κ? In previous publications [32, 33] the unsuccessful choice of the main parameters revealed only the torus-chaos transition corresponding to the lower part of the full map in Fig. 27. To check the robustness of the detected attractors we calculated the boundaries of the basic regimes and found that they changed on the map only in absolute coordinates but their relative positions remained the same. In particular, the set n=3.15, k=10 shifts the boundaries of the regimes towards smaller values of **α** but larger values of Q are then required to reveal all regimes (data not shown). A small detuning of Repressilators via the parameter **α** mismatches did not remove the discovered ways for chaos development and the appearance of hysteretic areas.

The formation of symmetric and asymmetric limit cycles inside quasiperiodic and chaotic regimes, as well as the generation of extended regions with chaos, is a reliable consequence of the specific choice of the QS equation used in system (1) for the coupling of the smooth ring oscillators. The other ODE model of coupled Repressilators used in [27] is different from system (1) by the location of the protein accompanying the production of autoinducer. However, the other type of multistability manifested as the coexistence of in-phase and anti-phase regular and complex oscillations  is generated in extended areas of control parameters [28].

We speculate that the QS-dependent multistability may be realized in other synthetic models with different versions of autoinducer production and its targets which may be important for the understanding of the origin of variability, not only in genetic networks.

**5. Acknowledgements**

This work is partially supported by grants Russian Foundation for Basic Research 15-02-03236 and 17-01-00070. We thank Nikolay Sukhorukov for the calculations of Lyapunov exponents.

# References

============================================================

[1] Pikovsky, A, Rosenblum, M, Kurths, J. Synchronization A Universal Concept in Nonlinear Sciences, Cambridge University Press; 2003.
[2] Pisarchik, AN, Feudel, U. Control of multistability, Physics Reports. 540 (2014) 167-218.
[3] Koseska, A, Volkov, E, Kurths, J. Oscillation quenching mechanisms: Amplitude vs. oscillation death, Physics Reports. 531 (2013) 173-99.
[4] Taylor, AF, Tinsley, MR, Showalter, K. Insights into collective cell behaviour from populations of coupled chemical oscillators, Physical Chemistry Chemical Physics. 17 (2015) 20047-55.
[5] Nagornov, R, Osipov, G, Komarov, M, Pikovsky, A, Shilnikov, A. Mixed-mode synchronization between two inhibitory neurons with post-inhibitory rebound, Communications in Nonlinear Science and Numerical Simulation. 36 (2016) 175-91.
[6] Prasad, A, Sakai, K, Hoshino, Y. Direct coupling: a possible strategy to control fruit production in alternate bearing, Scientific Reports. 7 (2017) 39890.
[7] Kuramoto, Y, Battogtokh, D. Coexistence of Coherence and Incoherence in Nonlocally Coupled Phase Oscillators, Nonlinear Phenomena in Complex Systems. 5 (2002) 380-5.
[8] Omel'chenko, OE, Wolfrum, M, Yanchuk, S, Maistrenko, YL, Sudakov, O. Stationary patterns of coherence and incoherence in two-dimensional arrays of non-locally-coupled phase oscillators, Physical Review E. 85 (2012) 036210.
[9] Tinsley, MR, Nkomo, S, Showalter, K. Chimera and phase-cluster states in populations of coupled chemical oscillators, Nat Phys. 8 (2012) 662-5.
[10] Martens, EA, Thutupalli, S, Fourrière, A, Hallatschek, O. Chimera states in mechanical oscillator networks, Proceedings of the National Academy of Sciences. 110 (2013) 10563-7.
[11] Panaggio, MJ, Abrams, DM. Chimera states: coexistence of coherence and incoherence in networks of coupled oscillators, Nonlinearity. 28 (2015) R67.
[12] Zakharova, A, Kapeller, M, Schöll, E. Chimera Death: Symmetry Breaking in Dynamical Networks, Physical Review Letters. 112 (2014) 154101.
[13] Purcell, O, Savery, NJ, Grierson, CS, di Bernardo, M. A comparative analysis of synthetic genetic oscillators, Journal of The Royal Society Interface. 7 (2010) 1503-24.
[14] O'Brien, EL, Van Itallie, E, Bennett, MR. Modeling synthetic gene oscillators, Mathematical Biosciences. 236 (2012) 1-15.
[15] Mandal, MK, Sarkar, BC. Ring oscillators: Characteristics and applications, Indian Journal of Pure and Applied Physics. 48 (2010) 136-45.
[16] Elowitz, MB, Leibler, S. A synthetic oscillatory network of transcriptional regulators, Nature. 403 (2000) 335-8.
[17] Niederholtmeyer, H, Sun, ZZ, Hori, Y, Yeung, E, Verpoorte, A, Murray, RM, Maerkl, SJ. Rapid cell-free forward engineering of novel genetic ring oscillators, eLife. 4 (2015) e09771.
[18] Potvin-Trottier, L, Lord, ND, Vinnicombe, G, Paulsson, J. Synchronous long-term oscillations in a synthetic gene circuit, Nature. 538 (2016) 514-7.
[19] Gao, XJ, Elowitz, MB. Synthetic biology: Precision timing in a cell, Nature. 538 (2016) 462-3.
[20] Hellen, EH, Volkov, E, Kurths, J, Dana, SK. An Electronic Analog of Synthetic Genetic Networks, PLoS ONE. 6 (2011) e23286.
[21] Hellen, EH, Kurths, J, Dana, SK. Electronic circuit analog of synthetic genetic networks: Revisited, Eur Phys J Special Topics. 226 (2017) 1811-28.
[22] Garg, N, Manchanda, G, Kumar, A. Bacterial quorum sensing: circuits and applications, Antonie van Leeuwenhoek. 105 (2014) 289-305.

[23] Smith, RP, Boudreau, L, You, L. Engineering Cell-to-Cell Communication to Explore Fundamental Questions in Ecology and Evolution. in: SJ Hagen, (Ed.). The Physical Basis of Bacterial Quorum Communication. Springer New York, New York, NY, 2015. pp. 227-47.

[24] Danino, T, Mondragon-Palomino, O, Tsimring, L, Hasty, J. A synchronized quorum of genetic clocks, Nature. 463 (2010) 326-30.

[25] McMillen, D, Kopell, N, Hasty, J, Collins, JJ. Synchronizing genetic relaxation oscillators by intercell signaling, Proceedings of the National Academy of Sciences. 99 (2002) 679-84.

[26] Kuznetsov, A, Kærn, M, Kopell, N. Synchrony in a Population of Hysteresis-Based Genetic Oscillators, SIAM Journal on Applied Mathematics. 65 (2004) 392-425.

[27] Garcia-Ojalvo, J, Elowitz, MB, Strogatz, SH. Modeling a synthetic multicellular clock: Repressilators coupled by quorum sensing, Proceedings of the National Academy of Sciences of the United States of America. 101 (2004) 10955-60.

[28] Potapov, I, Volkov, E, Kuznetsov, A. Dynamics of coupled repressilators: The role of mRNA kinetics and transcription cooperativity, Physical Review E. 83 (2011) 031901.

[29] Li, C, Chen, L, Aihara, K. Stochastic synchronization of genetic oscillator networks, BMC Systems Biology. 1 (2007) 6.

[30] Yi, QZ, Zhang, JJ, Yuan, ZJ, Zhou, TS. Collective dynamics of genetic oscillators with cell-to-cell communication: a study case of signal integration, The European Physical Journal B. 75 (2010) 365-72.

[31] Yi, Q, Zhou, T. Communication-induced multistability and multirhythmicity in a synthetic multicellular system, Physical Review E. 83 (2011) 051907.

[32] Ullner, E, Koseska, A, Kurths, J, Volkov, E, Kantz, H, García-Ojalvo, J. Multistability of synthetic genetic networks with repressive cell-to-cell communication, Physical Review E. 78 (2008) 031904.

[33] Ullner, E, Zaikin, A, Volkov, EI, García-Ojalvo, J. Multistability and Clustering in a Population of Synthetic Genetic Oscillators via Phase-Repulsive Cell-to-Cell Communication, Physical Review Letters. 99 (2007) 148103.

[34] Hellen, EH, Volkov, E. Flexible dynamics of two quorum-sensing coupled repressilators, Physical Review E. 95 (2017) 022408.

[35] Brewster, RC, Jones, DL, Phillips, R. Tuning Promoter Strength through RNA Polymerase Binding Site Design in Escherichia coli, PLOS Computational Biology. 8 (2012) e1002811.

[36] Davis, JH, Rubin, AJ, Sauer, RT. Design, construction and characterization of a set of insulated bacterial promoters, Nucleic Acids Research. 39 (2011) 1131-41.

[37] Hellen, EH, Dana, SK, Zhurov, B, Volkov, E. Electronic Implementation of a Repressilator with Quorum Sensing Feedback, PLoS ONE. 8 (2013) e62997.

[38] Ermentrout, B. Simulating, Analyzing, and Animating Dynamical Systems: A Guide to XPPAUT for Researchers and Students. Philadelphia, PA, SIAM; 2002.

[39] Doedel, EJ, Fairgrieve, TF, Sandstede, B, Champneys, AR, Kuznetsov, YA, Wang, X. AUTO-07P: Continuation and bifurcation software for ordinary differential equations. 2007.

[40] Taylor, AF, Tinsley, MR, Wang, F, Huang, Z, Showalter, K. Dynamical Quorum Sensing and Synchronization in Large Populations of Chemical Oscillators, Science. 323 (2009) 614-7.

[41] De Monte, S, d'Ovidio, F, Danø, S, Sørensen, PG. Dynamical quorum sensing: Population density encoded in cellular dynamics, Proceedings of the National Academy of Sciences. 104 (2007) 18377-81.

[42] Bondarenko, VE, Cymbalyuk, GS, Patel, G, DeWeerth, SP, Calabrese, RL. Bifurcation of synchronous oscillations into torus in a system of two reciprocally inhibitory silicon neurons: Experimental observation and modeling, Chaos: An Interdisciplinary Journal of Nonlinear Science. 14 (2004) 995-1003.

[43] Rosenblum, M, Pikovsky, A. Self-Organized Quasiperiodicity in Oscillator Ensembles with Global Nonlinear Coupling, Physical Review Letters. 98 (2007) 064101.

[44] Schreiber, I, Marek, M. Transition to chaos via two-torus in coupled reaction-diffusion cells, Physics Letters A. 91 (1982) 263-6.
[45] Sporns, O, Roth, S, Seelig, FF. Chaotic dynamics of two coupled biochemical oscillators, Physica D: Nonlinear Phenomena. 26 (1987) 215-24.
[46] Hakim, V, Rappel, W-J. Dynamics of the globally coupled complex Ginzburg-Landau equation, Physical Review A. 46 (1992) R7347-R50.
[47] Nakagawa, N, Kuramoto, Y. Collective Chaos in a Population of Globally Coupled Oscillators, Progress of Theoretical Physics. 89 (1993) 313-23.